\documentclass{PoS}

\usepackage{natbib}
\newcommand{\apj}{ApJ}
\newcommand{\mnras}{MNRAS}
\newcommand{\pasj}{PASJ}
\usepackage{url}

\title{Probing Very Early Stage of Radio Source Evolution in NGC 1275 with VERA}

\ShortTitle{Probing Very Early Stage of Radio Source Evolution in NGC 1275 with VERA}

\author{\speaker{Hikaru Chida,}$^{ab}$   Hiroshi Nagai,$^b$ Kazunori Akiyama,$^{bd}$ Motoki Kino,$^c$ Mareki Honma,$^b$ Kyoshi Nishijima $^a$ and GENJI programme collaboration\\
\llap{$^a$}Department of Physics, Tokai University, 4-1-1 Kitakaname, Hiratsuka-shi, Kanagawa 259-1292, Japan\\
\llap{$^c$}National Astronomical Observatory of Japan, 2-21-1 Osawa, Mitaka, Tokyo 181-8588, Japan\\
\llap{$^b$}Korea Astronomy and Space Science Institute, 776 Daedukdae-ro, Yusong, Daejon 305-348, Korea\\
\llap{$^d$}Department of Astronomy, Graduate School of Science, The University of Tokyo, 7-3-1 Hongo, Bunkyo-ku, Tokyo 113-0033, Japan\\
 E-mail: \email{hikaru.chida@nao.ac.jp}}

\FullConference{12th European VLBI Network Symposium and Users Meeting,\\
                 7-10 October 2014\\
                 Cagliari, Italy}

\abstract{
We report on VLBI obsevations of NGC 1275 with VLBI Expolaration of Radio Astrometry (VERA) and Very Long Baseline Array (VLBA) during 2008 and 2013 at 15, 22 and 43 GHz. Our observations provide long-term variations in the radio flux and structure of the inner jet in NGC 1275 on sub-parsec scales, particularly at C3 which is a new radio component accountable for the recent active state in radio regime since early-2000s. We found the apparent velocity of C3 was $\beta_{\rm{app}} = 0.267 \pm 0.007 $, which was consistent with typical velocities of the hot spot and/or radio lobes in the young radio sources. Furthermore, the radio flux and size of C3 had simultaneously increased during September 2007 and May 2012 with keeping optically-thin spectra, suggesting that the particle acceleration occurred inside C3 as generally seen in the hot spots of the radio lobes. Our new observations suggest that C3 is a new-born hot spot and/or radio lobe at a very early stage of its evolution. Our results imply the radio lobes in the radio galaxies might be already formed on sub-parsec scales from the central SMBH and might play an important role to generate $\rm \gamma$-ray emission in the young radio sources.}

\begin{document}
\section{Introduction}
The nearby radio galaxy NGC 1275 (3C 84, Per A) is ideally suited to probe the physical nature of the relativistic jet in connection with its sub-parsec-scale structure observed with VLBI thanks to its proximity ($z$=0.0176; 1 mas = 0.35 pc).
NGC 1275 hosts many lobe-like structures on various spatial scales, indicating that it underwent recurrent nuclear activities in the past.
In recent, its nucleus has been reactivated again.
The Fermi Gamma-ray Spase Telescope has detected GeV $\rm \gamma$-ray emission much brighter than its upper limit measured by the EGRET in 1990s, which was accompanied with enhanced radio activities \citep[e.g.][]{Abdo et al. 2009}.
High-resolution observations with the VLBI Exploration of Radio Astrometry (VERA) detected an ejection of a new radio component C3 on sub-parsec scales \citep{Nagai et al. 2010}, which was already emerged in 2002 \citep{Suzuki et al. 2012} and has been accountable for recent radio activities from 2000s to 2011 \citep[e.g.][]{Nagai et al. 2010, Suzuki et al. 2012}. 
The physical nature of C3 is quite important for understanding not only the recent activities in NGC 1275, but also how the radio lobes were formed in NGC 1275 in the past activities.
For probing the physical properties of C3, we have performed VERA and the Very Long Baseline Array (VLBA) observations of NGC 1275.

\section{Observations}
We have intensively monitored NGC 1275 at 22 GHz with VERA on the framework of {\it Gamma-ray Emitting Notable AGN Monitoring by Japanese VLBI} (GENJI) {\it Programme} \citep{Nagai et al. 2013}, with intervals of about 2 weeks since October 2010.
In each observation, 3C 84 has been observed at typically 4--6 scans with an exposure time of 5 minutes at different hour angles.  
Data were reduced Astronomical Image Processing System (AIPS) developed by National Radio Astronomy Observatory (NRAO) in the same way as \cite{Nagai et al. 2013}.
The source images were reconstructed with {\tt modelfit} and a self-calibration technique implemented in the DIFMAP software package.

We also combined results of previous VERA observations from September 2007 to October 2010 \citep{Nagai et al. 2010} for tracing long-term variations in the inner-jet structure on sub-parsec scales. 
Furthermore, we also reduced VLBA data at 15 GHz and 43 GHz performed by MOJAVE program \footnote{\url{http://www.physics.purdue.edu/MOJAVE/sourcepages/0316+413.shtml}} and Boston University blazar group \footnote{\url{http://www.bu.edu/blazars/VLBA_GLAST/0316.html}}, respectively, for measuring the radio spectrum of C3.


%

\section{Results}
\begin{figure}[t]
\begin{center}
\includegraphics[height=4.5cm]{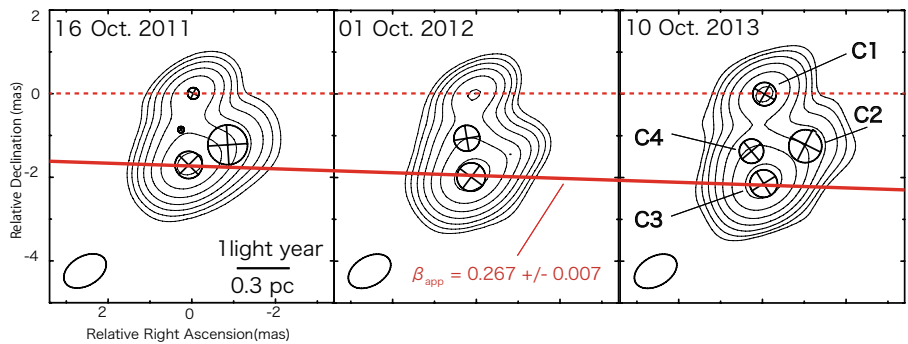}
\end{center}
\caption{
The inner jet of NGC 1275 on sub-parsec scales at 22 GHz traced by VERA observations at Jan. 2011, Oct. 2012 and Oct. 2013 with circular Gaussian components imposed. 
The images are restored with an elliptical Gaussian beam with FWHM sizes of 1.1 mas  $\times$ 0.7 mas at a position angle of $-60^\circ $.
 In order to compare with \cite{Nagai et al. 2010}, we unified the beam size to \cite{Nagai et al. 2010}.
The contours indicate levels of $3\sigma \times \sqrt{2^{n}}$ ($n = -1,0,1,2,3\cdots$), where $\sigma$ is the largest image noise of 73 mJy/beam among our VERA observations.
\label{fig:mas-scale image}
}
\end{figure}

The milliarcsecond-scale structure in NGC 1275 can be represented by four circular Gaussian components (Figure \ref{fig:mas-scale image}).
The jet extends toward the south from the northern core component C1.
A new radio component C3 was moving to the south from the core C1 during 2007 and 2013. 
Its motion is mostly along with the north-south direction, although it shows a slight motion in the east-west direction which was also reported during 2003 and 2008 \citep{Suzuki et al. 2012}.
The apparent velocity of C3 is estimated to be $\beta_{\rm{app}} = 0.267 \pm 0.007 $ assuming a rectilinear motion.

\begin{figure}[t]
\begin{center}
\includegraphics[height=6.4cm]{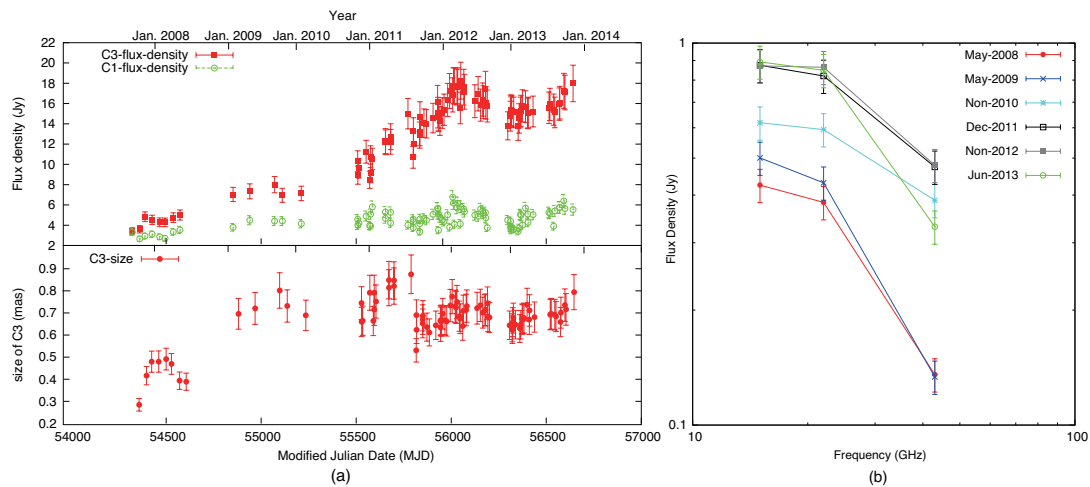}
\end{center}
\caption{
The time variation of the flux density, size and radio spectrum of C3. Errors are 1 $\sigma$.
{\it (a)}: The flux density of C1 and C3 (upper panel) and the size of C3 (lower panel) at 22 GHz as a function of Modified Julian Date (MJD) from 54358 to 56646.
{\it (b)}: The time variation in the radio spectrum of C3 between 15 GHz and 43 GHz from May 2008 to Jun 2013.
The radio flux of C3 at each frequency is measured from the image restored with a synthesized beam of VERA shown in Figure 1, by taking the total flux of a square area with a width of 0.47 mas centered on the position of C3. \label{fig:time variation of C3}
}
\end{figure}

We show time variations in the flux density and the FWHM size of C3 in Figure \ref{fig:time variation of C3} (a).
The flux variation at 22 GHz mostly originated in C3, because the flux densities of C1 and other components were stable through observing epochs. 
The flux density of C3 had monotonically increased until May 2012 and been relatively stable after.
Interestingly, the size of C3 follows a similar trend; it had increased until the August 2011 and been relatively constant as well.

Figure \ref{fig:time variation of C3} (b) shows the time variation in the radio spectrum of C3.
The total flux drastically varied at each frequency but keeping the shape of the spectrum.
The time-averaged spectral indexes are -0.23 $\pm$ 0.13 between 15 GHz and 22 GHz and -1.16 $\pm$ 0.09 between 22 GHz and 43 GHz.
It suggests that the synchrotron radiation from C3 is optically-{\it thin} at $> 22$ GHz and partially optically-{\it thin} at $\sim 22$ GHz.

\section{Discussion \& Summary}
The activities in C3 are unusual as jet components, whose fluxes generally decreased with time due to the adiabatic and/or synchrotron cooling effects. 
The drastic activities and observational properties in C3 can be explained by a scenario that C3 is a terminal component of the jet like the hot spot and the radio lobe.

First, our observations revealed that the apparent velocity of C3 was $\beta_{\rm{app}} = 0.267 \pm 0.007 $, which was consistent with typical velocities of the hot spot at the early stages 
(kinematic ages of $\leq$ kyr) of the radio lobe evolution (0.1 $\sim$ 0.3$c$). This is consistent with the scenario that C3 is the terminal component (lobe / hot spot) rather than a jet knot.

Second, our new observations indicate that the particle acceleration occurs in C3, which is one of the representative character of the hot spot that is the terminal shock of the jet.
As shown in Figure \ref{fig:time variation of C3} (a), the increment in the flux density and size simultaneously occurred in the period of September 2007 and August 2011. 
In general, such an expansion of radio components with the flux enhancement can be explained by the adiabatic expansion of optically{\it -thick} plasma.
Otherwise (i.e. in optically-{\it thin} case), it requires continuous injection of the relativistic particles attributed to the particle acceleration to maintain or increase the radio flux against the adiabatic cooling and synchrotron cooling effects.
Our results on the radio spectrum of C3 disfavor the expansion of the optically-thick plasma in C3, indicating that the particle acceleration is required to occur in C3 to keep or increase its radio flux.

Our new observations suggest that the new radio component C3 of NGC 1275 is most likely a radio lobe and/or a hot spot at a very early stage of radio lobe evolution.
This suggests that the radio lobes in radio galaxies might be already formed on sub-parsec scales at the vicinity of SMBHs as C3.
For the recent active state in $\rm \gamma$-ray regime, our results implies that the relativistic leptons can be also supplied from the particle-acceleration region inside C3,
which might play an important role to generate inverse-compton emission in $\rm \gamma$-ray regime.
We will work on further investigations on C3 in the context of the evolution of the young radio source and mechanism of $\rm \gamma$-ray emission in NGC 1275.

\end{document}